\documentclass[11pt,journal,draftcls,letterpaper,onecolumn]{IEEEtran}
\IEEEoverridecommandlockouts
\usepackage{cite}
\usepackage{amsmath,amssymb,amsfonts}
\usepackage{algorithmic}
\usepackage{graphicx}
\usepackage{textcomp}
\usepackage{subcaption}
\usepackage{balance}
\usepackage{physics}

\usepackage{xcolor}
\def\BibTeX{{\rm B\kern-.05em{\sc i\kern-.025em b}\kern-.08em
    T\kern-.1667em\lower.7ex\hbox{E}\kern-.125emX}}

\begin{document}

\title{Anomaly Detection Based on Generalized Gaussian Distribution approach for Ultra-Wideband (UWB) Indoor Positioning System}

\author{\IEEEauthorblockN{1\textsuperscript{st} Fuhu Che}
\IEEEauthorblockA{\textit{School of Computing and Engineering} \\
\textit{University of Huddersfield}\\
Huddersfield, UK \\
f.che@hud.ac.uk}\\
\and
\IEEEauthorblockN{2\textsuperscript{nd} Qasim Zeeshan Ahmed}
\IEEEauthorblockA{\textit{School of Computing and Engineering} \\
\textit{University of Huddersfield}\\
Huddersfield, UK \\
q.ahmed@hud.ac.uk}\\
\and
\IEEEauthorblockN{3\textsuperscript{rd} Faheem A. Khan}
\IEEEauthorblockA{\textit{School of Computing and Engineering} \\
\textit{University of Huddersfield}\\
Huddersfield, UK \\
f.khan@hud.ac.uk}\\
\and
\IEEEauthorblockN{4\textsuperscript{th} Pavlos I. Lazaridis}
\IEEEauthorblockA{\textit{School of Computing and Engineering} \\
\textit{University of Huddersfield}\\
Huddersfield, UK \\
p.lazaridis@hud.ac.uk}\\

}
\maketitle
\begin{abstract}
With the rapid development of the Internet of Things (IoT), Indoor Positioning System (IPS) has attracted significant interest in academic research. Ultra-Wideband (UWB) is an emerging technology that can be employed for IPS as it offers centimetre-level accuracy. However, the UWB system still faces several technical challenges in practice, one of which is Non-Line-of-Sight (NLoS) signal propagation. Several machine learning approaches have been applied for the NLoS component identification. However, when the data contains a very small amount of NLoS components it becomes very difficult for existing algorithms to classify them. This paper focuses on employing an anomaly detection approach based on Gaussian Distribution (GD) and Generalized Gaussian Distribution (GGD) algorithms to detect and identify the NLoS components. The simulation results indicate that the proposed approach can provide a robust NLoS component identification which improve the NLoS signal classification accuracy which results in significant improvement in UWB positioning system.

\end{abstract}

\section{Introduction}
Wireless communications is playing a major role in changing our life styles~\cite{Luo-2018, Yousif-2016,Alluhaibi-2020,Riaz-2020,Farooq-2019,Ahmed-2013a,Pan-2017,Nair-2016}.The Global Navigation Satellite System (GNSS) has brought tremendous convenience to human life~\cite{Park-2020}. Unfortunately, the GNSS signal does not work for an indoor environment because the satellite signal attenuates heavily while it passes through walls~\cite{Win-2018}. Ultrawide Bandwidth (UWB) system represents a more promising technique than other technologies for indoor high-accuracy localization since it can provide good time resolution and robust signalling~\cite{Guo-2020}. However, the UWB system faces numerous challenges in practical scenarios, including signal acquisition, multipath effects, Non-Line of Sight (NLoS) propagation and so on~\cite{Waqas-2021,Ahmed-2007}. NLoS propagation has been considered as one of the major challenges especially for high-precision UWB indoor positioning systems (IPS)~\cite{Rezazadeh-2019,Sang-2018}. In the NLoS environment, the direct signal between the transmitter and receiver is blocked or attenuated by obstacles, which causes additional signal propagation delay, resulting in inaccurate estimation of the distance. Therefore, the localization accuracy of the system is reduced significantly~\cite{Ahmed-2020}. 

Anomaly detection is a common application of machine learning (ML) algorithms, mainly used for unsupervised learning problems and some supervised learning problems~\cite{Piccinni-2019}. Different ML algorithms have been employed to address the NLoS ranging error by different researchers.  One of the early ML-based NLoS identification was conducted in~\cite{Alsindi-2009}. In this algorithm, the variance of Time of Arrival (ToA) was used to identify the NLoS component by setting an appropriate threshold. In~\cite{Marano-2010,Wymeersch-2012}, the authors proposed Least Square-Support Vector Machine (LS-SVM) algorithm to distinguish between the Line of Sight (LoS) and NLoS components. In this paper, they showed that the ML approach can be employed to improved the accuracy error caused by the NLoS components. Later, several ML algorithms combined with different features of the waveform UWB signal have been investigated for NLoS identification. In~\cite{Yan-2017}, different hypothesis tests for NLoS identification based on the amplitude and delay statistics of typical UWB channel impulse response (CIR) were conducted. In~\cite{Fan-2019}, an unsupervised ML approach based on Gaussian mixture models is proposed to identify the NLoS from the unlabelled data. Finally, in~\cite{Yang-2019}, the authors used the Relevance Vector Machine (RVM) to distinguish LoS/NLoS components based on the superior kernel machine for range measurement.

It is hard for these algorithms to classify the different types of anomalies~\cite{Che2-2020} despite showing an improved localization accuracy. In addition, if this training data which has different types of anomalies is employed, it becomes difficult for the algorithms to learn from the positive or negative instances. Therefore, the actual accuracy of the classification will be quite low. The objective of this paper is to classify these LoS and NLoS components with the help of Gaussian Distribution (GD) and Generalized Gaussian Distribution (GGD)~\cite{Ahmed-2013,Ahmed-2014,Ahmed2-2014,Ahmed2-2015}. These algorithms will help us classify the LoS and NLoS components more accurately especially in the case where the NLoS components are in a small amount. The advantages of GD and GDD are: 
\begin{itemize}
    \item Strict mathematical theory support, strong interpretability, thus simplifying the classification problems~\cite{Che2-2020};
    \item Be able to find key samples that are critical to the task~\cite{Che-2020};
    \item Highest prediction with the limit feature strategy.
\end{itemize} 
Furthermore, these algorithms will help when different types of abnormalities of the signal are present as it will improve the accuracy of the UWB positioning system.

This paper is organized as follows: Section 2 mainly explains the UWB localization system model. In Section 3, the problem formulation is expressed and details of our proposed algorithms is presented. Section 4  discusses the environment in which the data was collected, and the hardware used in the data collection. Moreover, the feature used NLoS classification is also presented in this section. Section 5 presents the performance evaluation of the proposed algorithms and discusses the results; The summarization of what it has been accomplished discusses in section 6.

\section{System Model}
\subsection{Transmitted Signal}
UWB is based on transmitting an extremely short pulse having a duration of nanoseconds~\cite{Dardari-2010}. The pulse has a very low power spectral density~\cite{Ahmed-2015,Hao-2007,Hu-2005}. The high bandwidth and extremely short pulses can reduce the effect of multipath interference which provides robust communication between the transmitter and receiver~\cite{Hao-2007,Hu-2005,Beaulieu-2008}. It has the capability to reach high accuracy within centimetres due to its characteristics. Therefore, UWB is one of the most suitable choice where the accuracy factor is required by the application. Considering the IR-UWB system is transmitted by $K$ pulses with a period of $T_{p}$  that consists of certain frames~\cite{Ahmed-2015}, the transmitted signal is shown as
\begin{equation}\label{eq-1}
s(t)=\sqrt{E}\sum_{k=1}^{K-1}{p(t- kT_p)},
\end{equation}
where $E$ is the energy of the single pulse, $k$ represents the index of the pulses, and $p(t)$ is the pulse waveform which in our case is the $2^{nd}$ derivative of a Gaussian pulse~\cite{Ahmed-2007,Beaulieu-2008}. 
\subsection{Received Signal}
The transmitted signal experiences a multipath channel and the received signal can be expressed as~\cite{Molisch-2003}
\begin{equation}\label{eq2}
r(t)=\sum_{n=1}^{N}{a_np(t-\tau_n)+n(t)},
\end{equation}
where $N$ is the number of received multipath compoments,$a_n$ and $\tau_n$ represent the amplitude and delay of the $n$-th path respectively, and $n(t)$ is the Additive White Gaussian Noise (AWGN) with zero mean and two-sided power spectral density $N_0/2$~\cite{Ryan-2017,Ahmed-2008,Ahmed2-2008}.
\section{Problem Formulation}
The UWB based IPS consists of two types of nodes. In one node type, the position is fixed and they are referred to as anchors. In another type, the position is unknown and is called as a tag. The position of this tag is measured with respect to the anchors. In UWB based IPS, the ToA algorithm is employed due to it exploits good time resolution and ease of implementation. To locate the tag, we need to estimate the distance $d$ between the tag and the anchors. When clear LoS exists between the anchors and tag, the estimate distance $\hat{d}$ can be calculated by 
\begin{equation}~\label{eq-3}
\hat{d}=c\times\tau, 
\end{equation}
where $\tau$ is propagation time and $c$ is the speed of the signal which propagates with the same speed of light. However, in the NLoS environment, the direct path between the anchors and tag is absent or delayed as the signal goes through the obstruction or other materials in the surroundings. In this case, the propagation time $\tau$ will be longer, which causes the positive bias and ranging error for the UWB IPS system. To combat this problem, in this paper, anomaly detection based on Gaussian Distribution (GD) and Generalized Gaussian Distribution (GGD) algorithms are proposed to classify the LoS and NLoS environment that will result in improving the accuracy of the UWB IPS. In order to evaluate the proposed algorithms, we compare the proposed algorithms with the existing Naïve Bayes (NB) algorithm. Let us briefly review the principle of the NB algorithm before introducing the proposed algorithms.
\subsection{Naïve Bayes, (NB)}
The NB algorithm can improve the classification accuracy by calculating each feature of the data~\cite{Che-2020}. The Bayesian principle is employed in this algorithm. According to the Bayesian principle, the given probability can be measured as
\begin{equation}\label{eq-4}
 P(l\mid x)=\frac{P(x\mid l)(P(l)}{P(x)}
\end{equation}
where$P(l \mid x)$ is the conditional probability of $l$ given $x$, $P(l)$ is the prior probability of the $l$, $P(x\mid l)$ is the probability of the $x$ given the condition and the $P(x)$ is the prior probability of the event $x$.

Given a dataset $D=\{{x_{1},x_{2},\cdots,x_{M}}\}$, the dataset $D$ contain a set of classes $\{l=0~\textrm{or}~1\}$, when applying the NB algorithm indicating the LoS $(l=0)$ or NLoS $(l=1)$ environment, respectively. The probability of the class $l$ under the attribute conditions $x_{1},x_{2},...,x_{M}$ can be calculated as 
\begin{equation}\label{eq-5}
P(l\mid x_{1},x_{2},\cdots, x_{M})=\frac { P(x_{1},x_{2},\cdots,x_{M}\mid l)P(l)}{P(x_{1},x_{2},\cdots, x_{M})} .
\end{equation}
As these feature $\{{x_{1},x_{2},\cdots,x_{M}}\}$ are independent from one another, therefore,
\begin{eqnarray}\label{eq-6}
P(x_{1},x_{2},\cdots, x_{M}\mid l) 
&=&P(x_{1}\mid l)\times \cdots \times P(x_{M}\mid l) \nonumber\\
&=&\prod_{i=1}^{M}P(x_{i}\mid l).
\end{eqnarray}
Substituting the above (\ref{eq-6}) into (\ref{eq-5}), we get 
\begin{eqnarray}\label{eq-7}
P(l\mid x_{1},x_{2},\cdots, x_{M}) 
&=&\frac {P(x_{1},x_{2},\cdots,x_{M}\mid l)P(l)}{ P(x_{1},x_{2},\cdots, x_{M})} \nonumber\\
&=&\frac{\prod_{i=1}^{M}P(x_{i}\mid l) P(l)}{\prod_{i=1}^{M}P(x_{i})}.
\end{eqnarray}
\subsection{Gaussian Distribution, (GD)}
The Probability Distribution Function (PDF) of Gaussian Distribution (GD) with variable $x$, mean $\mu$, and variance $\sigma^{2}$ can be written as ~\cite{book:alouini}
\begin{equation}\label{eq-8}
 P(x,\mu,\sigma ^{2})=\frac{1}{\sqrt{2\pi}\sigma}\exp\left(-\frac{(x-\mu)^2}{2\sigma ^2}\right).
\end{equation}
Given a data set $\{x_{1},x_2,\cdots,x_{M}\}$, $x\in {\mathbb{R}}^{n}$, and $x$ is Gaussian distributed, then in order to find the GD we just need to estimate the mean  $\hat{\mu}$ and variance $\hat{\sigma}^2$ which is calculated as 
\begin{eqnarray}\label{eq-9}
\hat{\mu} &=&\frac{1}{M-1}\sum_{i=1}^{M}{x_{i}}\nonumber\\
\hat{\sigma}^2&=&\frac{1}{M-1}\sum_{i=1}^{M}{(x_{i}-\hat{\mu})}^2 
\end{eqnarray}
%
\subsection{Generalized Gaussian Distribution, (GGD)}
In some cases, the GD model cannot identify the data abnormality clearly as there are two key parameters to be modelled which are the mean and variance. Therefore, instead of GD, Generalized Gaussian Distribution (GGD) can be adopted~\cite{Ahmed-2013,Ahmed2-2015}. The GGD, PDF is given as
\begin{equation}\label{eq-10}
P(x, \mu, \alpha, \beta)=\frac{\beta}{2 \alpha \Gamma(1/\beta)} \exp \left( -\frac{|x-\mu|}{\alpha} \right)^\beta
\end{equation}
where $\mu$ is the mean, $\beta$ determines the shape of the PDF, and $\Gamma(\cdot)$ is the gamma function. The variance $\sigma^2$ and the kurtosis $\kappa$ is given as~\cite{Ahmed-2014}
\begin{eqnarray}\label{eq-11}
\sigma^2 &=& \frac{\alpha^2 \Gamma(3/\beta)}{\Gamma(1/\beta)}\nonumber\\
\kappa &=& \frac{\Gamma(5/\beta)\Gamma(1/\beta)}{\Gamma(3/\beta)^2}-3.
\end{eqnarray}
Given a data set $\{x_{1},x_2,\cdots,x_{M}\}$, $x\in {\mathbb{R}}^{n}$, in order to find the GGD we need to estimate the mean  $\hat{\mu}$, variance $\hat{\sigma}^2$ and kurtosis $\hat{\kappa}$ which is calculated as 
\begin{eqnarray}\label{eq-12}
\hat{\mu} &=&\frac{1}{M-1}\sum_{i=1}^{M}{x_{i}}\nonumber\\
\hat{\sigma}^2&=&\frac{1}{M-1}\sum_{i=1}^{M}{(x_{i}-\hat{\mu})}^2 \nonumber\\
\hat{\kappa}&=&\frac{\frac{1}{M-1}\sum_{i=1}^{M}{(x_{i}-\hat{\mu})}^4}{\left[\frac{1}{M-1}\sum_{i=1}^{M}{(x_{i}-\hat{\mu})}^2\right]^2}-3,
\end{eqnarray}
where estimate of kutosis $\hat{\kappa}$ can be used to measure the shape parameter $\beta$ and estimate of variance $\hat{\sigma}^2$ can help determine the scale parameter $\alpha$ of the GGD.

\begin{table}[t]
\centering
\caption{Configurations of the MDEK 1001 Kit.}\label{table-1}
\begin{center}
\begin{tabular}{|l|l| }
\hline
\textbf{Properties} & \textbf{Values}  \\
\hline
Data Rate & 6.8 Mbps \\
\hline
Frequency & 3993.6 MHz \\
\hline
Bandwidth & 499.2 MHz\\
\hline
Channel & 2\\
\hline
PRF & 16 MHz\\
\hline
\end{tabular}
\end{center}
\end{table}

\section{Data Collection Process and Methodology}
In this paper, LoS and NLoS scenario is considered. In the LoS condition, there are no physical obstacles between the anchor and tag. While, in the NLoS case, there will be an obstacle (e.g., a wall or a person) between the anchor and tag so that the direct path does not exist. The main path of the signal will be attenuated or blocked by this obstacle, which will cause a propagation delay of the signal. The experiment is set up in a general room environment having an area of around $16 m^2$. We employ MDEK1001 kits as the UWB hardware. Four anchors and one tag will be set up in this experiment. The MDEK 1001 kits configurations used in this experimental evaluation are shown in Table~\ref{table-1} where $PRF$ stands for Pulse Repetition Frequency. All the work is carried out with the help of MATLAB (R2019b).

\subsection{LoS Measurement}
Four anchors are placed with the tag in a complete LoS condition. The actual distance between the anchor and the tags is measured with the help of an electronic tape measure and is averaged at least $5$ times to minimize the error caused by measurement. The tag is connected to the PC and the relevant dataset will be recorded by the PC in Tera term software. Total $500$ samples in LoS will be collected.
\subsection{NLoS Measurement}
For NLoS scenario data collection, an obstacle will be put between the anchor and tag to block the direct communication path. Total $50$ samples in NLoS condition are collected. The variance of the estimated distance and the power of the first path will be used to identity the NLoS environment in this paper. Range estimation method is based on the statistical characteristics of the estimated ToA distance. The variance of estimate distance can be used to identify the NLoS when the variance exceeds the set threshold. 
\subsection{Methodology}
The DECAWAVE investigates the NLoS identification based on the first path power threshold which has been widely used in different applications and system implementations currently~\cite{Che2-2020}. This approach is based on taking the power difference between the estimated received (RX) power and first-path power. The first-path power is measured in $dBm$ and is calculated as follows~\cite{Che-2020}.
\begin{equation}\label{eq-15}
\textrm{First Path Power Level}=10 \log_{10}\left(\frac{F_1^2+F_2^2+F_3^2}{N^2}\right)-A,
\end{equation}
where $F_1$, $F_2$, and $F_3$ represents the first, second and third harmonics of the first-path signal respectively. $A$ is fixed to $113.77$ for a PRF of $16$MHz, and $N$ represents the Preamble Accumulation Count value.

The received power of the signal in $dBm$ is calculated as follow 
\begin{equation}\label{eq-16}
\textrm{RX Power Level}=10 \log_{10}\left(\frac{CIR_P\times 2^{17}}{N^2}\right)-A,
\end{equation}
where $CIR_P$ is the power of the CIR of DW1000 chip. The received signal power is more in case of NLoS environment as compared to LoS environment due to higher number of multipath components in the former. The power in the first path for LoS signal is expected to be higher than NLoS signal. This difference between the received and first-path power can be used to identify the LoS and NLoS signal and is given as
\begin{equation}\label{eq-17}
\textrm{Difference in Power} = \textrm{RX Power Level} - \textrm{FP Power Level}.
\end{equation}
Let us now proceed to the simulation and evaluation of these proposed methods.
\section{Simulation and Evaluation}

\begin{figure}[t]
     \centering
             \includegraphics[width=1.0\linewidth]{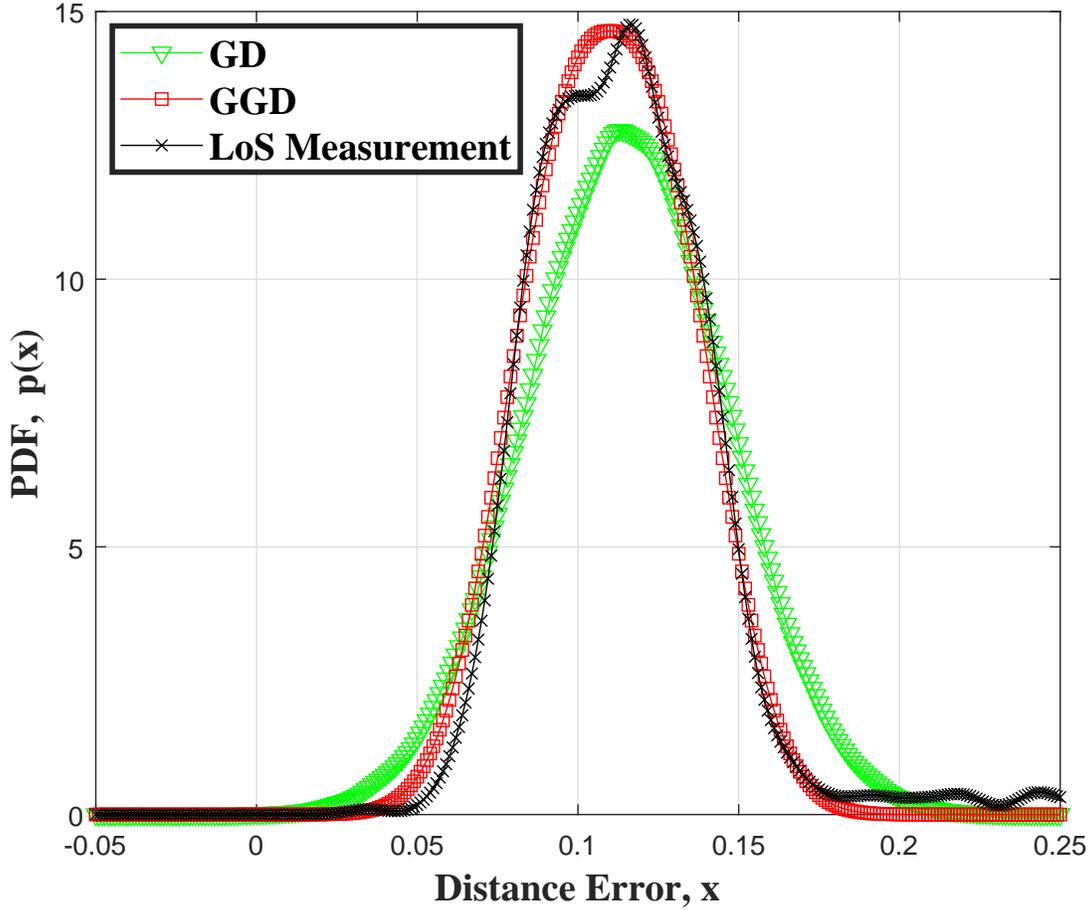}
         \caption{Line of Sight (LoS) error distance calculation.}
         \label{fig-2-LoS}
\end{figure}

\begin{figure}[t]
     \centering
             \includegraphics[width=1.0\linewidth]{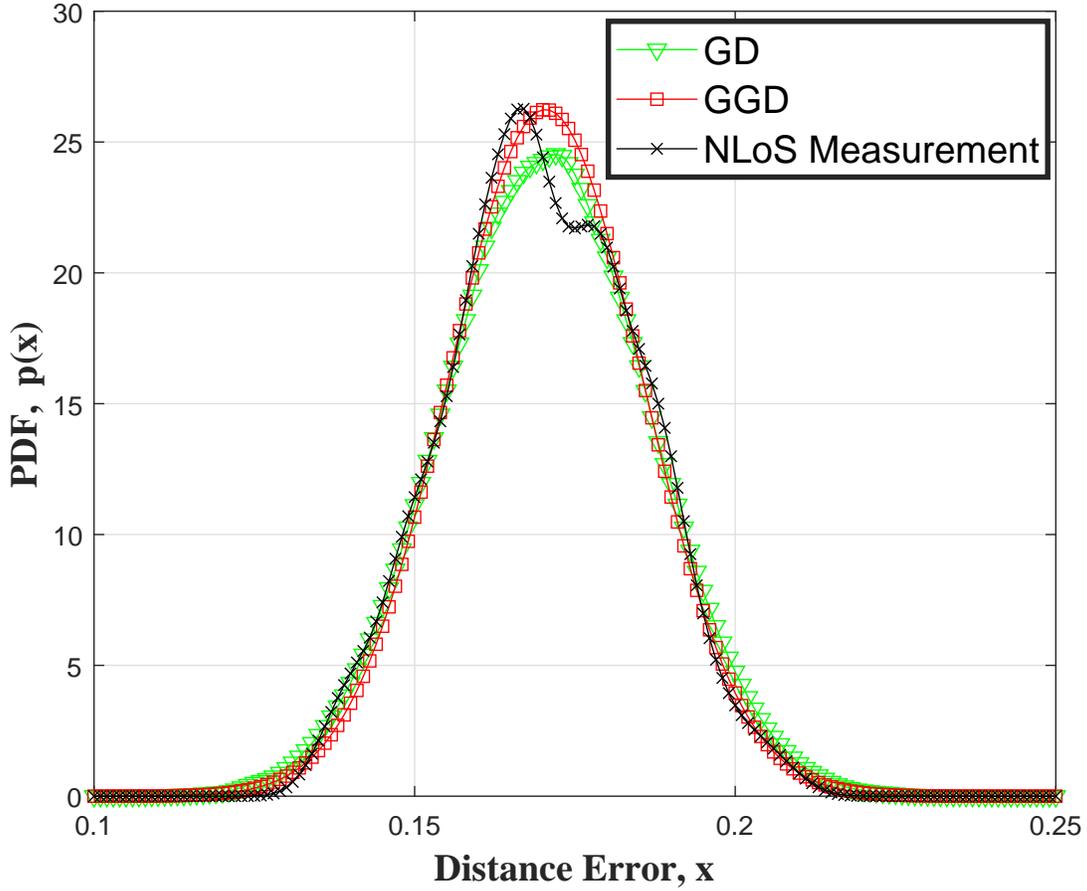}
         \caption{Non Line of Sight (NLoS) error distance calculation.}
         \label{fig-3-NLoS}
\end{figure}
In Figure~\ref{fig-2-LoS} and Figure~\ref{fig-3-NLoS} the probability distribution function (PDF) of LoS and NLoS are plotted. The PDF is plotted by measuring the  distance error which is calculated by using more than $2000$ reading. The distance error is simply the accurate distance measured minus the distance measured by the MDEK 1001 kit. It can also be observed from the figures that the distance error is less than $0.11$m for the LoS and $0.17$m for the NLoS signal. From both the figures it can be observed that the distance error can be more accurately evaluated with the help of Generalized Gaussian distribution (GGD) as compared to the Gaussian distribution (GD). This accurate pdf will further help us quantify the NLoS and LoS components accurately resulting in a significantly improved positioning system. 
\section{Conclusion}
In this paper, anomaly detection based on GD and GGD is proposed for an UWB based indoor localization system. These algorithms can improve the accuracy of the system by classifying the LoS and especially the NLoS components accurately. The variance of the estimated distance and the power of the first path are used as features to identity the NLoS environment. The simulation result shows that both GD and GGD algorithm has good anti-NLOS characteristics and can maintain good positioning accuracy as compared to NB. Furthermore, we have also shown that measured LoS and NLoS data is more close to GGD as compared to GD. In the future, we plan to design the optimal and receiver operating characteristics for GGD for UWB based indoor localization system.



\balance

\vspace{12pt}


\begin{thebibliography}{00}

\bibitem{Luo-2018}
Y. Luo, T. Ratnarajah, J. Xue and F. A. Khan, ``Interference Alignment in Two-Tier Randomly Distributed Heterogeneous Wireless Networks Using Stochastic Geometry Approach," IEEE Systems Journal, vol. 12, no. 3, pp. 2238-2249, Sept. 2018. 

 
\bibitem{Yousif-2016}
E. H. G. Yousif, M. C. Filippou, F. Khan, T. Ratnarajah and M. Sellathurai, ``A new LSA-based approach for spectral coexistence of MIMO radar and wireless communications systems," 2016 IEEE International Conference on Communications (ICC), 2016. 

\bibitem{Alluhaibi-2020}
O. Alluhaibi, Q. Z. Ahmed, E. Kampert, M. D. Higgins and J. Wang, ``Revisiting the Energy-Efficient Hybrid D-A Precoding and Combining Design for mm-Wave Systems," in IEEE Transactions on Green Communications and Networking, vol. 4, no. 2, pp. 340-354, June 2020.

\bibitem{Riaz-2020}
S. Riaz, F. A. Khan, S. Saleem and Q. Z. Ahmed, ``Reducing the Mutual Outage Probability of Cooperative Non-Orthogonal Multiple Access," in IEEE Transactions on Vehicular Technology, vol. 69, no. 12, pp. 16207-16212, Dec. 2020.

\bibitem{Farooq-2019}
A. Farooq, Q. Z. Ahmed and T. Alade, ``Indoor Two Way Ranging using mm-Wave for Future Wireless Networks," 2019 Proceedings of Emerging Technologies Conference (EMT), 2019, pp. 1.   

\bibitem{Ahmed-2013a}
Q. Z. Ahmed, K. Park, M. Alouini and S. Aïssa, "Optimal linear detectors for nonorthogonal amplify-and-forward protocol," 2013 IEEE International Conference on Communications (ICC), 2013, pp. 4829-4833.

\bibitem{Pan-2017}
Y. Pan, C. Pan, H. Zhu, Q. Z. Ahmed, M. Chen and J. Wang, "Content offloading via D2D communications based on user interests and sharing willingness," 2017 IEEE International Conference on Communications (ICC), 2017, pp. 1-6.

\bibitem{Nair-2016}
M. Nair, Q. Z. Ahmed and H. Zhu, ``Hybrid Digital-to-Analog Beamforming for Millimeter-Wave Systems with High User Density," 2016 IEEE Global Communications Conference (GLOBECOM), 2016, pp. 1-6.

\bibitem{Park-2020}
H. Park, A. Camps, J. Castellvi and J. Muro, ``Generic Performance Simulator of Spaceborne GNSS-Reflectometer for Land Applications," in IEEE Journal of Selected Topics in Applied Earth Observations and Remote Sensing, vol. 13, pp. 3179--3191, 2020.
\bibitem{Win-2018}
M. Z. Win, Y. Shen and W. Dai, ``A Theoretical Foundation of Network Localization and Navigation," in Proceedings of the IEEE, vol. 106, no. 7, pp. 1136--1165, July 2018.

\bibitem{Guo-2020}
X. Guo, N. R. Elikplim, N. Ansari, L. Li and L. Wang, "Robust WiFi Localization by Fusing Derivative Fingerprints of RSS and Multiple Classifiers," in IEEE Transactions on Industrial Informatics, vol. 16, no. 5, pp. 3177-3186, May 2020.


\bibitem{Waqas-2021}
W. B. Abbas, F. Che, Q. Z. Ahmed, F. A. Khan and T. Alade, ``Device Free Detection in Impulse Radio Ultrawide Bandwidth Systems," Sensors, vol. 21, no. 9, pp. 1--19, Sept. 2021.


\bibitem{Ahmed-2007}
Q. Z. Ahmed and L. Yang, ``Performance of Hybrid Direct-Sequence Time-Hopping Ultrawide Bandwidth Systems in Nakagami-$m$ Fading Channels," 2007 IEEE 18th International Symposium on Personal, Indoor and Mobile Radio Communications, 2007, pp. 1-5.

\bibitem{Rezazadeh-2019}
J. Rezazadeh, M. Moradi, K. Sandrasegaran and R. Farahbakhsh, ``Transmission Power Adjustment Scheme for Mobile Beacon-Assisted Sensor Localization," in IEEE Transactions on Industrial Informatics, vol. 15, no. 5, pp. 2859-2869, May 2019.

\bibitem{Sang-2018}
C. L. Sang, M. Adams, T. Hörmann, M. Hesse, M. Porrmann and U. Rückert, ``An Analytical Study of Time of Flight Error Estimation in Two-Way Ranging Methods," 2018 International Conference on Indoor Positioning and Indoor Navigation (IPIN), Nantes, 2018, pp. 1-8.

\bibitem{Ahmed-2020}
Q. Z. Ahmed, M. Hafeez, F. A. Khan and P. Lazaridis, ``Towards Beyond 5G Future Wireless Networks with focus towards Indoor Localization" 2020 IEEE Eighth International Conference on Communications and Networking (ComNet), 2020, pp. 1-5.

\bibitem{Piccinni-2019}
G. Piccinni, F. Torelli and G. Avitabile, ``Distance Estimation Algorithm for Wireless Localization Systems Based on Lyapunov Sensitivity Theory," in IEEE Access, vol. 7, pp. 158338-158348, 2019.

\bibitem{Alsindi-2009}
N. Alsindi, C. Duan, Jinyun Zhang and Tsutomu Tsuboi, ``NLOS channel identification and mitigation in Ultra Wideband ToA-based Wireless Sensor Networks," 2009 6th Workshop on Positioning, Navigation and Communication, Hannover, 2009, pp. 59-66.

\bibitem{Marano-2010}
S. Maranò, W. M. Gifford, H. Wymeersch and M. Z. Win, ``NLOS identification and mitigation for localization based on UWB experimental data," in IEEE Journal on Selected Areas in Communications, vol. 28, no. 7, pp. 1026-1035, September 2010.

\bibitem{Wymeersch-2012}
H. Wymeersch, S. Marano, W. M. Gifford and M. Z. Win, ``A Machine Learning Approach to Ranging Error Mitigation for UWB Localization," in IEEE Transactions on Communications, vol. 60, no. 6, pp. 1719-1728, June 2012.

\bibitem{Yan-2017}
L. Yan, Y. Lu and Y. Zhang, ``An Improved NLOS Identification and Mitigation Approach for Target Tracking in Wireless Sensor Networks," in IEEE Access, vol. 5, pp. 2798-2807, 2017.

\bibitem{Fan-2019}
J. Fan and A. S. Awan, ``Non-Line-of-Sight Identification Based on Unsupervised Machine Learning in Ultra Wideband Systems," in IEEE Access, vol. 7, pp. 32464-32471, 2019.

\bibitem{Yang-2019}
X. Yang, ``NLOS Mitigation for UWB Localization Based on Sparse Pseudo-Input Gaussian Process," in IEEE Sensors Journal, vol. 18, no. 10, pp. 4311-4316, 15 May15, 2018.

\bibitem{Che2-2020}
F. Che, A. Ahmed, Q. Z. Ahmed, S. A. R. Zaidi and M. Z. Shakir, ``Machine Learning Based Approach for Indoor Localization Using Ultra-Wide Bandwidth (UWB) System for Industrial Internet of Things (IIoT)," 2020 International Conference on UK-China Emerging Technologies (UCET), Glasgow, United Kingdom, 2020, pp. 1-4.

\bibitem{Ahmed-2013}
Q. Z. Ahmed, M. Alouini and S. Aissa, ``Bit Error-Rate Minimizing Detector for Amplify-and-Forward Relaying Systems Using Generalized Gaussian Kernel," in IEEE Signal Processing Letters, vol. 20, no. 1, pp. 55-58, Jan. 2013.


\bibitem{Ahmed2-2015}
Q. Z. Ahmed, K. Park and M. Alouini, ``Ultrawide Bandwidth Receiver Based on a Multivariate Generalized Gaussian Distribution," in IEEE Transactions on Wireless Communications, vol. 14, no. 4, pp. 1800-1810, April 2015.


\bibitem{Ahmed2-2014}
Q. Z. Ahmed, K. Park, M. Alouini and S. Aïssa, ``Compression and Combining Based on Channel Shortening and Reduced-Rank Techniques for Cooperative Wireless Sensor Networks," in IEEE Transactions on Vehicular Technology, vol. 63, no. 1, pp. 72-81, Jan. 2014

\bibitem{Ahmed-2014}
Q. Z. Ahmed, K. Park, M. Alouini and S. Aissa, ``Linear Transceiver Design for Nonorthogonal Amplify-and-Forward Protocol Using a Bit Error Rate Criterion," in IEEE Transactions on Wireless Communications, vol. 13, no. 4, pp. 1844-1853, April 2014.

\bibitem{Che-2020}
F. Che, A. Ahmed., Q. Z. Ahmed, and M. Z. Shakir,``Artificial intelligence for localisation of ultra-wide bandwidth (UWB) sensor nodes, " in AI for Emerging Verticals: Human-Robot Computing, Sensing and Networking (2020 ed.). IET



\bibitem{Dardari-2010}
D. Dardari, R. D'Errico, C. Roblin, A. Sibille and M. Z. Win, ``Ultrawide Bandwidth RFID: The Next Generation?," in Proceedings of the IEEE, vol. 98, no. 9, pp. 1570-1582, Sept. 2010.

\bibitem{Ahmed-2015}
Q. Z. Ahmed, S. Ahmed, M. Alouini and S. Aïssa, ``Minimizing the Symbol-Error-Rate for Amplify-and-Forward Relaying Systems Using Evolutionary Algorithms," in IEEE Transactions on Communications, vol. 63, no. 2, pp. 390-400, Feb. 2015.

\bibitem{Hao-2007}
K.~Hao and J.~A.~Gubner, ``The distribution of sums of path gains in the IEEE 802.15.3a UWB channel model," in IEEE Transactions on Wireless Communications, vol.~6, pp.~811-–816, March 2007.

\bibitem{Hu-2005}
B.~Hu and N.~C.~Beaulieu, ``Pulse shapes for ultrawideband communication systems," in IEEE Transactions on Wireless Communications, vol.~4, pp.~1789–-1797, July~2005.

\bibitem{Beaulieu-2008}
N.~C.~Beaulieu and B.~Hu, ``On determining a best pulse shape for multiple access ultra-wideband communication systems," in IEEE Transactions on Wireless Communications, vol.~7,
pp.~3589–-3596, September~2008.

\bibitem{Molisch-2003}
A.~F.~Molisch, J.~R.~ Foerster, and M.~Pendergrass, ``Channel models for ultrawideband personal area networks," in IEEE Wireless Communications, vol.~10, pp.~14–-21, December 2003.

\bibitem{Ryan-2017}
R. Husbands, Q. Z. Ahmed and J. Wang, ``Transmit antenna selection for massive MIMO: A knapsack problem formulation," 2017 IEEE International Conference on Communications (ICC), 2017, pp. 1-6.

\bibitem{Ahmed-2008}
Q. Z. Ahmed, W. Liu and L. Yang, ``Least Mean Square Aided Adaptive Detection in Hybrid Direct-Sequence Time-Hopping Ultrawide Bandwidth Systems," VTC Spring 2008 - IEEE Vehicular Technology Conference, 2008, pp. 1062-1066.

\bibitem{Ahmed2-2008}
Q. Z. Ahmed and L. Yang, ``Normalised Least Mean-Square Aided Decision-Directed Adaptive Detection in Hybrid Direct-Sequence Time-Hopping UWB Systems," 2008 IEEE 68th Vehicular Technology Conference, 2008, pp. 1-5.


\bibitem{book:alouini}
M.~K.~Simon and M.-S.~Alouini, {\em {D}igital {C}ommunication over {F}ading {C}hannels}.\newblock Wiley Series in Telecommunications and Signal Processing, 2005.






\end{thebibliography}
\end{document}